# The Chinese Equity Bubble: Ready to Burst

K. Bastiaensen‡, P. Cauwels‡, D. Sornette*,** R. Woodard* and W.-X. Zhou ***

July 10, 2009

*Amid the current financial crisis, there has been one equity index beating all others: the Shanghai Composite. Our analysis of this main Chinese equity index shows clear signatures of a bubble build up and we go on to predict its most likely crash date: July 17-27, 2009 (20%/80% quantile confidence interval).*

In the light of a global downturn a wide range of stimuli packages and fiscal gifts have been launched by many governments of which China makes no exception. However, in China these actions might have lead to an unsustainable rise in asset prices, a so-called "bubble". The Shanghai Composite is the best performing large stock market in 2009 and is up 65 per cent for the year, and rising. To reach a targeted GDP growth of 8%, Chinese policy has turned to a bank model of massive lending, which has provided China with sufficient liquidity to fuel this bubble. Talk of a China bubble is heard in the market, but when or even if it will collapse is unknown, as usual. Macroeconomic and other qualitative factors have been used as arguments for or against the hypothesis of a China asset bubble.

We take a different approach here by closely examining the price changes of the index. This work is based on Didier Sornette's group (in particular with Anders Johansen from 1995 to 2002, with Wei-Xing Zhou since 2002 and, since 2008, with the FCO group at ETH Zurich: www.er.ethz.ch/fco/) who pioneered studies to identify bubbles and obtain estimates on likely "crash dates", when rapid acceleration in asset prices becomes unsustainable, leading to a crash. Sornette's group applied their theories to correctly identify bubbles and ranges of crash dates for the US Housing boom[1] and the oil bubble in 2008[2], amongst others.

## Patterns emerging from complexity

Due to their very nature, financial markets exist of interacting players that are connected in a network structure. These interacting players, often referred to as interacting agents, are continuously influencing each other. In scientific literature it is said that such a system is subject to non-linear dynamics. Modeling such a system in full detail is practically impossible to do. That is why the long-term behavior of the global economy or the weather is quite hard to predict.

Recent research, however, has provided new tools to analyze complex non-linear systems without having to go through the simulation of all underlying interactions. When interacting agents are playing in a hierarchical network structure very specific emerging patterns arise.

Let us clarify this with an example[3]. After a concert the audience expresses its appreciation with applause. In the beginning, everybody is handclapping according to their own rhythm. The sound is like random noise. There is no imminence of collective behavior. This can be compared to financial markets operating in a steady-state where prices follow a random walk. All of a sudden something curious happens. All randomness disappears; the audience organizes itself in a synchronized regular beat, each pair of hands is clapping in unison. There is no master of ceremony at play. This collective behaviour emanates endogenously. It is a pattern arising from the underlying interactions. This can be compared to a crash. There is a steady build-up of tension in the system (like with an earthquake or a sand pile) and without any exogenous trigger a massive failure of the system occurs. There is no need for big news events for a crash to happen.

---

‡ The first two authors are employees of BNP Paribas Fortis, Fortis nv/sa. This article is the result of a personal project in which the authors express their personal view and it does not necessarily correspond to the view of BNP Paribas Fortis.
* ETH Zurich, Department of Management, Technology and Economics, Kreuzplatz 5, CH-8032 Zurich, Switzerland
** Swiss Finance Institute, c/o University of Geneva, 40 blvd. Du Pont d'Arve, CH 1211 Geneva 4, Switzerland
***School of Business, East China University of Science and Technology, Shanghai 200237, China

[1] Zhou W.-X. and Sornette D., Is there a real-estate bubble in the US?, Physica A, 361, 297-308, Feb. 2006.
[2] Sornette D., Woodard R. and Zhou W.-X., The 2006-2008 Oil Bubble: evidence of speculation, and prediction, Physica A 388, 1571-1576 (2009) (arXiv:0806.1170, June 2008).
[3] After Philip Ball, Critical Mass: How One Thing Leads to Another, 2004, Heinemann.

Financial markets can be classified as open, non-linear and complex systems[4]. They also exhibit emanating patterns as a result of which the "invisible hand" can be very shaky.

More then 40 years ago Benoit Mandelbrot[5] described the fractal structure of cotton prices and the emanating properties of fat tails and volatility clustering and Hyman Minsky proposed a theory for endogenous speculative bubble formation. More recently Robert Shiller and Alan Greenspan made the irrational exuberance paradigm fashionable[6,7]. These all fit in the framework of Complexity Economics, which describes the properties that emerge from interacting agents. It has become clear that herding behaviour in financial markets results in positive or negative feedback mechanisms causing price accelerations or decelerations and (anti)-bubble formation, where asset prices become detached from the underlying fundamentals

During the last decade, physicist-turned-financial economist Didier Sornette[8,9] and his group applied techniques used to model earthquakes, global warming, demographic patterns and the failure of materials to the financial markets. In a wide range of scientific publications, they found evidence of a simple and general theory of how, why and when stock markets crash[10].

## Results

The model leads to a log periodic power law (LPPL), whose formula and interpretation are thoroughly explained in references[9] and in a recent refinement of the theory[11]. It must be noted that a good fit of the model to the data series is not a 100% certainty for a crash, but it clearly points at a bubble formation. A critical point leads to a change in dynamics. Here the crash is most likely, but there exist a small yet finite possibility that the bubble will deflate more gently.

The result of the analysis is summarized below in Figure 1. We analyzed the Shanghai SSE Composite Index time series between October 15, 2008 and July 9, 2009. We increased the starting date of the LPPL analysis in steps of 15 days while keeping the ending date fixed, resulting in 10 fits. The figure shows observations of the SSEC Index as black dots (joined by straight lines) and the LPPL fits as smooth lines until the last day of analysis. The y-axis is logarithmically scaled, so that an exponential function would appear as a straight line and a power law function with a finite-time singularity would appear with a slightly upward curvature. Note that the LPPL fits to the observations exhibit this slightly upward curvature. The vertical and horizontal dashed lines indicate the date and price of the highest price observed, July 6, 2009. Extrapolations of the fits to 100 days beyond July 9, 2009 are shown as lighter dashed lines. The darker shaded box with diagonal hatching indicates the 20%/80% quantiles of the projected crash dates, July 17-27, 2009. The lighter shaded box with horizontal hatching indicates the range of all 10 projected crash dates, July 10 - August 10, 2009. These two shaded boxes indicate the most probable times (with the associated confidence levels) to expect peak and possible subsequent crash of the Index. The parameters of the fit confirm the faster-than-exponential growth of the Shanghai SSE Composite Index over this time interval, a clear diagnostic of the presence of a bubble.

---

[4] Eric Beinhocker, The Origin of Wealth, 2006, Harvard Business Press.
[5] Benoit Mandelbrot, The (Mis)Behaviour of Markets: A Fractal View of Risk, Ruin and Reward, 2004, Basic Books.
[6] http://www.irrationalexuberance.com
[7] Robert Shiller, Irrational Exuberance, 2001, Broadway Books.
[8] http://www.er.ethz.ch/people/sornette
[9] Didier Sornette, Why Stock Markets Crash: Critical Events in Complex Financial Systems, 2003, Princeton University Press.
[10] Sornette D. and Johansen A., Large financial crashes, Physica A 245, 411-422 (1997)
[11] Lin L., Ren R.E,Sornette D., A Consistent Model of `Explosive' Financial Bubbles With Mean-Reversing Residuals, arXiv:0905.0128, May'09

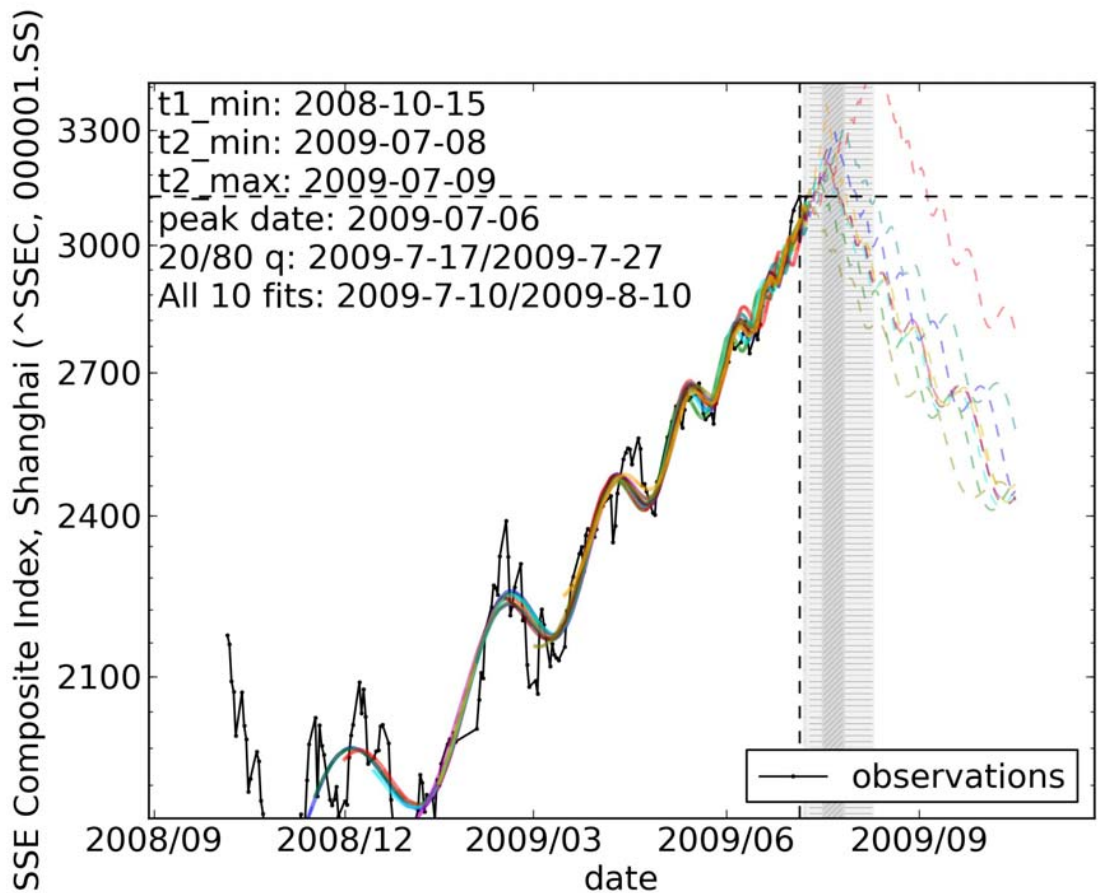

Figure 1: Shanghai Composite Index with LPPL result.

It must be noted that the model gives no indication on what happens after the critical point. It tells us the rise will end, but that might be with a crash or a slow decay.

## Conclusion

By the very nature of the model, this result gives us two conclusions. Firstly, there exists a bubble in the Shanghai Composite Index. Secondly, it will reach a critical level around July 17-27, 2009. This will lead to a change in regime which may be a crash or a more gently bubble deflation. An extended version of this note, with a careful assessment of the confidence intervals and comparisons with the previous Chinese bubble ending in Oct. 2007, will be released soon.